\documentclass{sig-alternate-10pt}
\usepackage[noadjust]{cite}
\usepackage{graphicx}
\usepackage{url}
\usepackage{pifont}
\usepackage{color}
\usepackage{soul}
 
\newcommand{\beq}{\begin{equation}}
\newcommand{\eeq}{\end{equation}}

\def\bearn{\begin{eqnarray*}}
\def\eearn{\end{eqnarray*}}
\def\barr{\begin{array}}
\def\earr{\end{array}}

\newcommand{\urlsamefont}[1]
{
\urlstyle{same}\url{#1}
}

\begin{document}

\title{I  Know Where You are and What You are Sharing:\\
  \bigskip \Large \bf \textsf Exploiting P2P Communications to Invade
  Users' Privacy}

\numberofauthors{1} \author{
  \alignauthor Stevens Le Blond\thanks{The first author contributed to this work while a Ph.D. student at INRIA.} \ Chao Zhang$^\dagger$ Arnaud Legout$^{\ddagger}$ Keith Ross$^\dagger$ Walid Dabbous$^{\ddagger}$\\
  \affaddr{$^*$ MPI-SWS, Germany \quad $^\dagger$ NYU-Poly, USA \quad
    $^\ddagger$ INRIA, France }}

\maketitle
\sloppy

\begin{abstract}

  In this paper, we show how to exploit real-time communication
  applications to determine the IP address of a targeted user.  We
  focus our study on Skype, although other real-time communication
  applications may have similar privacy issues.  We first design a
  scheme that calls an identified-targeted user
  \textit{inconspicuously} to find his IP address, which can be done
  even if he is behind a NAT.  By calling the user periodically, we
  can then observe the mobility of the user.  We show how to scale the
  scheme to observe the mobility patterns of tens of thousands of
  users.  We also consider the linkability threat, in which the
  identified user is linked to his Internet usage. We illustrate this
  threat by combining Skype and BitTorrent to show that it is possible
  to determine the filesharing usage of identified users.  We devise a
  scheme based on the identification field of the IP datagrams to
  verify with high accuracy whether the identified user is
  participating in specific torrents.  We conclude that any Internet
  user can leverage Skype, and potentially other real-time
  communication systems, to observe the mobility and filesharing usage
  of tens of millions of identified users.

\end{abstract}

\section{Introduction}
\label{sec:intro}

The cellular service providers are capable of tracking and logging our
whereabouts as long as our cell phones are powered on.  Because the
web sites we visit see our source IP addresses and cookies, the web
sites we frequently visit -- such as Google \cite{google} and Facebook
\cite{facebookweb} -- can also track our whereabouts to some extent. 
Although tracking our whereabouts can be considered a
major infringement on our privacy, most people are not terribly
concerned, largely because they trust that the cellular and major
Internet application providers will not disclose this information.
Moreover, these large companies have privacy policies, in which they
assure their users that they will not make location history, and other
personal information, publicly available.

In this paper, we are not concerned about whether large brand-name
companies can track our mobility, but instead about whether smaller
less-trustworthy entities can leverage the Internet to {\em
  periodically track our whereabouts}.  Is it possible, for example,
for an ordinary user with modest financial resources, operating from
his or her home, to periodically determine the IP address of a
targeted and identified Internet user and to link it to this user's
Internet activities (e.g., file sharing)?  We will show that the
answer to this question is yes!

Real-time communication (e.g., VoIP and Video-over-IP) is enormously
popular in the Internet today. As shown in Table \ref{tab:intro-apps},
the applications Skype, QQ, MSN Live, and Google Talk together have
more than 1.6 billion registered users.

Real-time communication in the Internet is naturally done peer-to-peer
(P2P), i.e., datagrams flow directly between the two conversing users.
The P2P nature of such a service, however, exposes the IP addresses of
all the participants in a conversation to each other. Specifically, if
Alice knows Bob's VoIP ID, she can establish a call with Bob and
obtain his current IP address by simply sniffing the datagrams
arriving to her computer. She can also use geo-localization services
to map Bob's IP address to a location and ISP.  If Bob is mobile, she
can call him periodically to observe his mobility over, say, a week or
month.  Furthermore, once she knows Bob's IP address, she can crawl
P2P file-sharing systems to see if that IP address is
uploading/downloading any files.  Thus VoIP can {\em potentially} be
used to collect a targeted user's location. And VoIP can {\em
  potentially} be combined with P2P file sharing to determine what a
user is uploading/downloading.  This would clearly be a serious
infringement on privacy.

\begin{table}
   \centering
  \begin{tabular}{|c|c|c|c|}
    \hline
    App         & \# Users & Dir & P2P\\
    \hline
    Skype       & 560M & \ding{51} & \ding{51}\\
    MSN Live    & 550M & \ding{55} & \ding{51}\\
    QQ          & 500M & \ding{51} & \ding{51}\\
    Google Talk & 150M & \ding{55} & \ding{51}\\
    \hline
  \end{tabular}
\caption{Number of users claimed by Skype
    \cite{skypenumbers}, MSN Live \cite{MSNnumbers}, QQ
    \cite{QQnumbers}, and Google Talk \cite{gtalknumbers} and for each
    of these systems, whether it has a directory service and employs
    P2P communications.}
\label{tab:intro-apps}
\end{table}

However, for such a scheme to be effective, there are several technical
challenges:

\begin{itemize}

\item For a specific targeted individual -- such as Bob Smith,
  28 years old, living in Kaiserslautern Germany --
  can Alice determine with certainty his VoIP ID?

\item Can Alice determine which packets come from Bob (and thereby
  obtain his IP address)? Indeed, during call setup, Alice may receive
  packets from many other peers. In addition, can Alice call Bob
  inconspicuously, so that Alice can periodically call Bob and get his
  IP address without Bob knowing it? Finally, can Alice obtain Bob's
  address, even when Bob configures his VoIP client to block calls
  from Alice?

\item If Bob's IP address, found with VoIP, is the same as an IP
  address found in a P2P file-sharing system, then we cannot conclude
  with certainty that Bob is downloading the corresponding file, since
  Bob may be behind a NAT (with the matching IP address being the
  public IP address of the NAT). Thus, is it possible to verify that
  Bob is indeed uploading/downloading the files, given that NATs are
  widely deployed in the Internet?

\end{itemize}

In this paper, using Skype, we develop a measurement scheme to meet
all the above challenges.  (This may be possible with other VoIP
systems as well, which we leave for future work.) Our main
contributions are the following:

\begin{itemize}

\item \textit{We develop a scheme to find a targeted person's Skype ID
    and to inconspicuously call this person to find his IP address,
    even if he is behind a NAT.}  By carefully studying Skype packet
  patterns for a Skype caller, we are able to distinguish packets
  received from the Skype callee from packets received from many other
  peers.  Having identified these packets, we extract the callee's IP
  address from the headers of the packets.  Furthermore, through
  experimentation, we determine how to obtain the IP address of the
  callee fully inconspicuously, that is, without ringing or notifying
  the user.  Finally, we show that Skype privacy settings fail to
  protect against our scheme.

\item \textit{We show our scheme can be used periodically to observe
    the mobility of Skype users.}  By scaling our scheme, we
  demonstrate that Skype does not implement counter measures to hinder
  such schemes. Although there are several challenges to measure the
  mobility of a large number of users, we show that it can be done
  efficiently and effectively.

\item \textit{We show that the scheme introduces a linkability threat
    where the identity of a person can be associated to his Internet
    usage.}  We illustrate this threat by combining Skype and
  BitTorrent to show that it is possible to determine the file-sharing
  usage of identified users.  One of the challenges here is that a
  BitTorrent user is often NATed, so that he may share his IP address
  with many other users.  When a common IP address is discovered in
  both Skype and BitTorrent, we immediately launch a verification
  procedure in which we simultaneously call the corresponding user and
  perform a BitTorrent handshake to the IP address, port and infohash
  (which identifies the file being shared).  We then use the
  identification field of the IP datagrams to verify with high
  accuracy whether an identified user is participating in specific
  torrents. To the best of our knowledge, we are the first ones to
  show that such a scheme can be used in the wild.

\end{itemize}

In addition to the technical contributions of this paper, another
contribution is that we are alerting Internet users (and the Skype
company as discussed in the next section) of a major privacy
vulnerability, whereby targeted users can have their mobility and
Internet usage tracked.  As of May 2011 (more than six months after
having notified the Skype company), all the schemes presented in this
paper are still valid.  We provide some relatively simple solutions so
that future real-time communication systems can be made less
vulnerable to these attacks.

One solution that would go a long way is to design the VoIP system so
that the callee's IP address is not revealed until the user accepts
the call. With this property, Alice would not be able to
inconspicuously call Bob. Moreover, if Alice is a stranger (that is,
not on Bob's contact list), and Bob configures his client to not
accept calls from strangers, then this design would prevent any
stranger from tracking him, conspicuously or otherwise.  However, even
with this solution in place, any friend of Bob, say Susan, can still
call him conspicuously and obtain his IP address. Susan could be Bob's
spouse, parent, employer, or employee, for example. It would be hard
for Susan to periodically track Bob this way, but Susan could still
(i) get Bob's current location, and (ii) check to see if Bob is
downloading content from a P2P file-sharing system.  Preventing these
attacks would require more fundamental changes in the VoIP system
(specifically, using relays by default) or more fundamental changes in
the underlying Internet protocols.

This paper is organized as follows.  We discuss the legal and ethical
considerations of this paper in Section~\ref{sec:legal}.  In
Section~\ref{sec:mapping}, we describe our scheme to determine the
current IP address of a person using Skype.  We then show that this
scheme can be used periodically to observe the mobility and
file-sharing usage of identified users in Section~\ref{sec:location}
and~\ref{sec:filesharing}.  Finally, we discuss some simple defenses
in Section~\ref{sec:defenses}, the related work in
Section~\ref{sec:related}, and we conclude in
Section~\ref{sec:conclusion}.

\section{Legal and Ethical Considerations}
\label{sec:legal}

In this measurement study, all testing involving identified users has
been performed on a small sample of volunteers who gave us their
informed consent to make measurements and publish results.
Unfortunately, the informed consent process for privacy, as for fraud
\cite{JakobSP08}, may significantly bias user behavior.  For example,
informed users may stop using Skype or BitTorrent.  For this reason,
we also needed to consider a larger sample of (anonymized) users in
order to accurately assess the amount of personal information that is
revealed by a normal usage, e.g., the mobility and file-sharing usage
of Skype users.  For the sake of privacy, we only stored and processed
anonymized mobility and file-sharing information.

Based on these arguments, the INRIA IRB approved this study.  In the
following, we describe our motivation to run privacy measurements, the
tests that we ran with volunteers, and the remaining measurements.

\paragraph{Motivation for Running Privacy Measurements} Internet users
publish a lot of personal information that can be exploited in
non-trivial ways to invade their privacy.  Indeed, recent research
demonstrates that personal information can be correlated in ways that
would have been hard to anticipate \cite{kirdaSP10}.  One goal of this
study is to show that any Internet user can leverage popular real-time
communication applications to observe the mobility patterns and
file-sharing usage of tens of millions of Internet users.  It is
important to give public visibility to these privacy issues, as they
constitute serious invasions into users' privacy, and can potentially
be used for blackmail and phishing attacks.

\paragraph{Volunteers} In this study, we have relied on two sets of
volunteers for which we have obtained informed consent.  The first set
comprises 14 research faculty in the CSE department at NYU-Poly for
which we have attempted to find the Skype IDs.

The second set comprises 20 people spread throughout the world (4 in
Asia, 2 in Australia, 7 in Europe, and 7 in USA) in cable and DSL
ISPs, with 10 users directly connectable and 10 users behind NAT.  We
deliberately chose users located in different continents and with
different Internet connectivity to observe a large diversity of user
and client behaviors.  We have relied on the second set of volunteers
to $(i)$ determine Skype packet patterns between caller and callee,
$(ii)$ develop and test inconspicuous calling, and $(iii)$ evaluate
the accuracy of mobility measurements.  After manual testing, we
called each volunteer 100 times and systematically observed one of the
three packet patterns described in Section~\ref{sec:mapping} between
caller and callee.  We also observed that our inconspicuous calling
procedure never notified them about the calls in any way.

\paragraph{Anonymized users} We relied on two samples of users for
which we did not store their personal information in this study.  We
first used a sample of 10,000 random users to quantify their mobility.
We then used a second sample of 100,000 random users that we used to
illustrate a linkability threat, where the identity of a person can be
associated to his Internet usage (e.g., file sharing).

We always collected the IP addresses of the anonymized users using
inconspicuous calls, which we validated on the volunteers.  Therefore,
\textit{no human contact was ever made with any of the anonymized
  users.}  Moreover, we processed and stored only anonymized
information, e.g., we anonymized all localization information,
downloaded content, and we did not store the IP addresses.  Details of
all anonymized information are given in Section~\ref{sec:location} and
\ref{sec:filesharing}.

\paragraph{Other considerations} In order to conform to the
responsible disclosure process, we informed the Skype company of our
conclusions in November 2010.  In addition, we did not perform any
reverse engineering on Skype binaries.  Finally, our measurements
generated at most $2.7$ calls per second and a few kilobytes of
bandwidth per second, so the load that we created on the Skype
infrastructure was marginal.

\section{Mapping a Person to an IP Address}
\label{sec:mapping}

In the following, we first describe how to find a targeted person's
Skype ID, that is a unique user ID of a person in Skype.  Then, we
present our scheme to find, based on a Skype ID, the IP address used
by this person.  We explain how to make this scheme inconspicuous for
the user, and we show that the privacy settings in Skype fail to
protect against our scheme.

\subsection{Finding a Person's ID}
\label{sec:mapping-ID}

When creating a Skype account, a user needs to provide an e-mail
address and Skype ID.  The user is also invited to provide personal
information, such as birth name, location, gender, age, and/or
website.  This information is recorded in the Skype directory.
Therefore, in attempting to define a person's Skype ID, the obvious
first step is to input into the directory's search service the
person's e-mail address or birth name.

When searching for a birth name, Skype will often return many results.
Along with these results, there is often side information, such as
city and country of residence.  As we will discuss below, if there is
still ambiguity about which Skype ID corresponds to the targeted
person, we can, using the methodology described in the following
section, inconspicuously call each of the candidate Skype IDs, obtain
a current or recent IP address for each of those IDs, and from the IP
addresses determine current city and ISP (which might be a University
or an employer ISP).  Such a procedure often determines a person's
Skype ID without ambiguity.  We briefly remark that if this search was
instead based on a service that doesn't provide a directory (such as
MSN Live or Google Talk), one may still be able to determine the ID by
scraping homepages, scraping pages from various social networks, or
simply by guessing.

To illustrate that one can easily find the Skype IDs for a set of
identified individuals, we attempt to find the IDs of the 14 research
faculty in the CSE department at NYU-Poly, all of whom gave us their
informed consent. By searching the corresponding 14 professional
e-mail addresses, we found 2 Skype IDs and by searching the
corresponding 14 birth names, we found 7 additional IDs with a single
match.  Among the 5 people for which we did not find a conclusive
Skype ID, there was multiple matching IDs for 4 and no matching Skype
ID only for 1.  For the professors with multiple candidate IDs, it
would have been possible to inconspicuously call each of the candidate
IDs (as described below), geo-localize each candidate, and most likely
pinpoint the correct ID.  In summary, among 14 NYU-Poly faculty
members, we found the Skype IDs for nine of them, and we could have
very possibly determined the IDs for four more.

\subsection{Finding a Person's IP Address}

We have seen how to find the Skype ID of a targeted person.  We now
discuss how, given the person's Skype ID, we can find the IP address
of the machine on which that person is currently active.  (If the
machine is behind a NAT, then we instead obtain the public IP address
of the NAT.)  The basic idea is to call the Skype ID, receive IP
datagrams from the machine on which that ID is currently logged in,
and sniff the packets to get the machine's IP address from the IP
header. We describe in the following when this IP address is available.

When the caller calls a Skype user who is currently off-line, the
Skype application will still provide to the caller the user's most
recent IP address, as long as the user was running Skype in the past
72 hours. For this reason, we are able to retrieve the IP address of a
Skype user that used Skype within the past 72 hours.

By examining traffic patterns to and from a Skype client when our
client makes a call to a Skype ID that has been active in the past 72
hours, we have observed that Skype behaves as follows.  At the time of
the call, the user may be in one of three possible states $(i)$ the
user is online and not behind a NAT; $(ii)$ the user is online and
behind a NAT; $(iii)$ the user is offline, but was online (with or
without a NAT) within the past 72 hours.  (There is also the
possibility that the user is logged in at more than one address
simultaneously.  We will discuss that case subsequently.)  For case
$(i)$, when the user (callee) is online and not behind a NAT, the
caller will initiate communication with the callee, sending packets
directly to the callee (with the callee's IP address in the
destination address field of the datagrams).  For case $(ii)$, when
the callee is online but behind a NAT, the callee will be instructed
(via the callee's supernode) to initiate communication to the caller.
In this case, the callee's public IP address will be in the source
address field of the incoming datagrams.  For case $(iii)$, when the
targeted user is offline (but was online in the past 72 hours), the
caller's Skype client will still attempt to call the targeted user,
using the IP address that was most recently observed by Skype in the
past 72 hours.  (If the targeted user is behind a NAT, the caller will
try to initiate a call, using the public IP address of the NATed
user.)  In this last case, the callee's most recent (public) IP
address can be determined from the IP datagrams. Thus, the callee's IP
address (current or most recent) can be extracted from the source and
destination fields of IP datagrams.

However, there is a major complication here.  In the process of
establishing a call, the call triggers communication with tens of IP
addresses (supernodes and relays).  As supernodes and relays are
hosted by Skype users, their IP addresses belong to a multitude of
address ranges that we cannot just filter out.  So it is complex to
determine which Skype datagrams are for direct communication with the
callee.  As Skype uses a proprietary protocol and encrypts the
payloads of its messages, we cannot perform direct packet inspection
to find packets originating from the callee.  To solve this problem,
we designed a scheme that relies solely on the packet patterns between
the caller and the various Skype nodes it is communicating with.

\begin{figure}[!t]
  \centering
  \includegraphics[width=1\columnwidth]{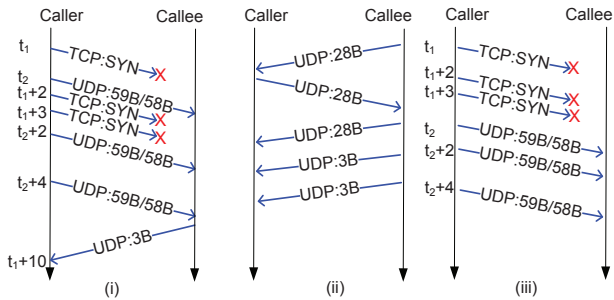}
  \caption{Communication pattern: (i) callee is online and public;
    (ii) callee online and behind a NAT; (iii) callee is
    offline. Crosses correspond to SYN packets that we dropped in
    order to call inconspicuously.}
  \label{fig:packet_pattern}
\end{figure}

To understand Skype's traffic, we placed calls to the second set of
volunteers for which we knew the IP addresses of (see
Section~\ref{sec:legal}).  We observed three identifiable patterns of
communication that take place between the caller and callee during the
call establishment phase.  By exploiting these patterns, we were able
to filter out the noise, such as communication with
supernodes. Fig.~\ref{fig:packet_pattern} shows these three patterns.

We observe the first pattern when the callee is online and public
(case $(i)$).  In that case, the caller will try to initiate the TCP
connection by sending a SYN packet.  We will see in
Section~\ref{sec:single-incons}, that we need to drop SYN packets to
make inconspicuous calls.  When the TCP timeout occurs, the caller
retransmits the SYN, making two tries after the initial attempt before
giving up.  The first timeout interval is 3 seconds and the second is
1 second.  In addition to the TCP packets, there are UDP packets
between the caller and the callee.  We always observe three 59 byte or
58 byte packets from caller to callee, and the intervals between them
are 2 seconds and 4 seconds. Thus, between caller and callee there is
a specific traffic pattern, which is shown in
Fig.~\ref{fig:packet_pattern} $(i)$.  There is also communication
between caller and supernodes; however, the communication with
non-callees does not exhibit the pattern in
Fig.~\ref{fig:packet_pattern} $(i)$.  In summary, by identifying the
IP address that has packets with the pattern in
Fig.~\ref{fig:packet_pattern} $(i)$, we identify the IP address of the
callee.  We remark that the TCP packets and UDP packets don't always
appear sequentially. Most of the time, they are mixed.

The second pattern is observed when the callee is online but behind a
NAT (case $(ii)$), that is, the caller cannot initiate communication
with the callee.  In that case, we have observed that the callee will
send a 28 byte UDP packet to the caller.  The caller replies with the
same size UDP packet.  Next, the caller and callee will exchange UDP
packets of varying sizes.  After about 10 seconds, the callee sends 3
byte UDP packets to the caller.  We do not observe these 3 byte UDP
packets from any other source besides the callee.  The pattern is
shown in Fig.~\ref{fig:packet_pattern} $(ii)$.

The last pattern occurs when the callee is offline but has been online
in the past 72 hours. In that case, the caller still attempts to call
the user at its last-seen IP address.  The pattern is shown in
Fig.~\ref{fig:packet_pattern} $(iii)$. Note this pattern has the same
structure as that of case $(i)$ except now there is no response from
the callee, since it is offline.

To make things even more complicated, a Skype ID can be simultaneously
online at more than one machine.  In this case, for each online
machine either the pattern in Fig.~\ref{fig:packet_pattern} $(i)$ or
$(ii)$ will occur once for each online machine.  We developed a script
that searches for the various patterns and identifies the callee's IP
address(es).

\subsection{Inconspicuous calling}
\label{sec:single-incons}

In the following, we define the tracking client as the Skype client we
use to exchange packets with a callee.  The tracking client is an
actual Skype client controlled by a script via the Skype API.
Importantly, each of the tracking client is not behind a NAT and,
therefore, has a public IP address.  Therefore, communication between
each tracking client and any user (NATed or not) will always be P2P
rather than relayed.

Whenever a Skype call comes in, it is accompanied with a ring and a
pop-up window for notification.  The callee then chooses to accept,
reject, or ignore the call.  (We use the terminology ``user'' and
``callee'' interchangeably, depending on context.)  Since the tracking
client actually makes calls to callees, if not designed carefully, it
will cause ringing and pop ups on the callees' machines.  Not only
would this disturb the callee, but it would expose the attacker.  We
therefore need to design our scheme so the tracking client exchanges
packets directly with the callee -- {\em without notifying the callee
  of the call}.

In our testing, we have observed that during call establishment, both
TCP and UDP packets are sent between the tracking client and the
callee.  We have found that if we prevent TCP connections from being
established with the callee, the callee will not be notified about the
call.  Thus, a possible simple solution is to have the tracking client
drop all TCP SYN packets sent to and from the callee.  However, at the
time when we make the call, we have no clue about the callee's IP
address, and we cannot tell whether an observed TCP SYN is going to
(or coming from) a Skype infrastructure node, a supernode, a relay
node, or the targeted callee.

To solve this problem, during each call, we prevent the establishment
of any new TCP connection by dropping all outgoing and incoming SYN
packets (to all IP addresses).  Note this procedure does not terminate
the tracking client's TCP connections that were in progress before
making the call (for example, an ongoing connection to a supernode).
With this simple mechanism, the callee is never notified, even if the
callee is behind a NAT.  To check that no pop ups appear, we tested
this scheme on the volunteers as described in Section~\ref{sec:legal}.

\subsection{Skype Privacy Settings}
\label{sec:skype-priv-settings}
Skype has two privacy settings to block calls from specific
people. The first setting, \textit{allows call from people in my
  Contact list only}, is a white list. The second setting called
\emph{blocked people} is a black list blocking all people whose Skype
ID is in this list.

We tested the impact of both settings on our scheme to inconspicuously
get the IP address of a callee. For the first setting, the caller was
not in the contact list of the callee. For the second setting, the
callee explicitly blocked the Skype ID of the caller. In both cases,
we were able to inconspicuously retrieve the IP address of the
callee. {\em In summary, we observed that Skype privacy settings fail
  to protect against our scheme.}

\section{Mobility of Skype Users}
\label{sec:location}

In the previous section, we presented a scheme to map a person's name
to an IP address.  We now investigate whether our scheme can be used
 to periodically observe the mobility patterns of large sets of Internet
users.

\subsection{Mobility of a Volunteer}

\subsubsection{Geo-Localize Skype Users}
\label{sec:single-geo}

In the following, we use MaxMind \cite{maxmind} to geo-localize the IP
addresses that are obtained from the tracking client, hence providing
us with the location of users.  MaxMind is a service that, given an IP
address, provides a city, country, and AS.  To determine city and
country, it first aggregates known IP locations from websites that ask
their users to provide their geographic location. Then, it uses
various heuristics to interpolate the location of other IP addresses.
MaxMind claims that it achieves $99.8$\% accuracy at the country level
and $83$\% on a city level for the US within a radius of 25 miles.

Apart from our set of volunteers, for the sake of user privacy, we
anonymized (using a salted hash) all location information.  Therefore,
we can tell when users change locations at the city, AS or country
scale, but not where they actually are.

\subsubsection{Example}
\label{sec:locatio-track-stevens}

\begin{figure}[!t]
\centering
\includegraphics[width=0.962\columnwidth]{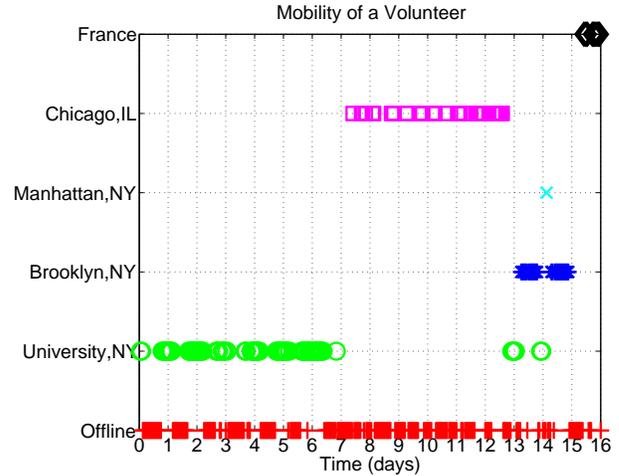}
\caption{Example of mobility of a volunteer.}
\label{fig:location-examples}
\end{figure}

To give a concrete idea of the kind of mobility that can be observed,
we plot in Fig.~\ref{fig:location-examples} the mobility of a user in
our second set of volunteers. (This volunteer has seen the paper and
has given us his consent for all the information about him disclosed.)
This person makes publicly available his birth name, gender, date of
birth, language, and city of residence in Skype.  By searching his
birth name and city on Facebook and LinkedIn, we are able to determine
his profession and employer.

We now briefly describe the mobility of this user. He confirmed to us
that during our measurement period he was first visiting a university
in New York; he then took a vacation in Chicago; then returned to
university and lodged in Brooklyn; and finally returned to his home in
France. Fig.~\ref{fig:location-examples} gives an accurate description
of the real mobility of this user during the measurement period. The
Manhattan location corresponds to an Internet cafe (confirmed by the
user).

We remark that if we had followed the mobility of the Facebook friends
of this user as well, we likely would have determined who he was
visiting and when.  In conclusion, mobility combined with information
from social networks can provide a vivid picture of the daily
activities of a targeted user. It is, in our opinion, a major privacy
concern for users of real-time communication systems.

Whereas this volunteer has an active mobility pattern well suited for
our illustrative purpose, a legitimate question is whether it is
possible to observe mobility for any Skype user.  We answer this
question in the following.

\subsection{Mobility of the Anonymized Users}

We now describe how to scale our scheme to measure the availability
and mobility of a representative sample of anonymized Skype users.  To
confirm the frequent mobility of Skype users, these users indeed need
to be often running Skype and from several locations.  In addition, we
are also interested in evaluating the cost of scaling our scheme and
in examining whether Skype employs counter measures to hinder it.

For the sake of privacy, we anonymized (as described in
section~\ref{sec:single-geo}) all location information, and we do not
store IP addresses. Therefore, we can only report aggregated
statistics, and not detailed user location information.

\subsubsection{Obtaining Millions of Skype IDs}
\label{sec:location-ID}

In the following, we show that one can easily retrieve a large number
of Skype IDs along with the personal information associated with these
IDs.  To this end, we use the Skype API to collect the IDs. For each
ID, we check whether the birth name and other personal information is
available. We do not store this information, but instead just note
whether it is available in the Skype user's profile.

The Skype public API provides a mechanism for third party applications
to control a Skype client.  This API operates as follows.  After
registering with the Skype client, the application can send to the
client plain text commands such as {\it search} and {\it call}.  The
Skype client then returns plain text messages to the application.  In
particular, the Skype API has a {\it search users} command that takes
a search string as a parameter and returns a list of users whose ID,
birth name, or e-mail address matches the string.  If the search
string contains {\it @}, the search is performed by e-mail address and
has to be an exact match.  If the search string is a valid Skype ID,
the search is performed on the birth name and ID.  Otherwise, the
search is made on the birth name only.  In addition to the Skype ID of
a user, this command will return any other personal information that
the user provided at registration, such as birth name, age, gender,
homepage, country, language, and other identifying information.

To build our search strings, we use a set of $580$K birth names that
we collected on Facebook using a similar technique as the one
described by Tang et al. \cite{facebook}.  This set is made up of
$66$K first names and $156$K last names.  We then combine the birth
names, first names, and last names, to obtain $802$K unique search
strings.  For each of these search strings, we send the {\it search
  users} command, which typically returns a long list of users, some
of whom didn't specify birth names.  We then aggregated these lists
together and obtained $13$M Skype IDs together with which identifying
information was available in the profile.  For these $13$M Skype IDs,
$88$\% provide their birth names and $82$\% provide either age,
gender, homepage, country or language identifying information (we only
store a binary information indicating whether a user has provided a
given personal information).  We note that even though we used
Facebook to build our search strings, we could use any database of
first and last names.

\subsubsection{Parallel Calling}
\label{sec:location-sequential}

From the Skype IDs obtained in the previous section, we select 100,000
Skype IDs at random.  From these 100,000 IDs, we then determine (using
the techniques discussed in Section~\ref{sec:mapping}) that 10,000
Skype IDs (10\%) have been active in the past 72 hours.  Finally, we
call these 10,000 Skype IDs on an hourly basis.  From this result
based on a random sample of 100,000 Skype users, we can extrapolate
that we can retrieve the IP address of approximately 10\% of all Skype
users at any time, which represents 56 million of users at any moment
in time.  We now describe the methodology to call the 10,000 Skype
IDs.

We deploy several tracking clients in parallel, each of which calls a
subset of the 10,000 Skype IDs.  The tracking client calls
sequentially all the Skype IDs in its subset, and then repeats the
procedure every hour.  We determine the IP address of each called
Skype ID using the inconspicuous call methodology described in
Section~\ref{sec:single-incons}. Based on this IP address we compute
the anonymized location of the user as described in
section~\ref{sec:single-geo}.

Scaling our scheme is challenging.  To be able to call 10,000 users on
an hourly basis, we need to deploy many tracking clients in parallel,
with each one sequentially making one call after another.  In order to
keep the number of parallel tracking clients to a reasonable level,
the time $s$ between two successive calls for a given client should be
short.

Indeed, there is an important tradeoff in considering an appropriate
value for $s$.  Consider that the tracking client calls one user,
waits $s$ seconds, terminates the call, and then repeats the process
with another user. If $s$ is large, our tracking client will call
users at a relatively low rate.  If $s$ is too small, we may terminate
the call before the packet pattern is initiated, in which case we may
incorrectly assign the IP address of the subsequent Skype ID to the
current Skype ID.  Thus, special care must be taken to associate the
IP addresses with the correct Skype IDs.

The simplest approach is, before making the subsequent call, to wait
long enough so that the complete packet pattern elapses.  Normally,
this takes about 15 seconds from when the first packet is observed
until the whole packet pattern occurs.  But if we wait 15 seconds
between each call, only 4 Skype IDs per minute can be probed.

To increase the calling rate, we performed further tests and observed
that $(a)$ once a packet pattern starts, it completes even if the call
is terminated before completion; $(b)$ all packet patterns begin
within three seconds after making the call.  Based on these
observations, by waiting three seconds before calling a new Skype ID,
we always see the pattern beginning before the end of the three second
interval, and also see the pattern complete (extending beyond the 3
seconds).  To verify claim (a), we randomly pick 500 users from our
Skype ID pool, and call them using two different values of $s$: 3
seconds and 20 seconds.  After comparing the mappings generated from
the two approaches, we observe that they are identical for all 500
random Skype users.  This implies that the interval of 3 seconds is
sufficiently large; we therefore use $s=3$ seconds in our
measurements.

To validate the accuracy of our scalable calling scheme, every 100
calls, we call a random Skype ID among our second set of volunteers
(see Section~\ref{sec:legal}).  We stress that these volunteers were
not in the contact list of the tracking clients, so the patterns
generated when calling them are identical to those of the other 10,000
users we are calling.  On the 1,368 calls that we made when volunteers
were online, we observed only 4 false positives ($0.3$\%) due to
patterns that have been reordered during parallel calling.  By
assigning each IP address to the only Skype user that is the most
often designated by the packet patterns, we were able to remove all
false positives.

\subsubsection{Cost of the Scaling}
\label{sec:cost-attack}

To call 10,000 users on a hourly basis, we run our tracking clients on
30 physical machines, each one with a different IP address.  Each
physical machine runs one Skype client and can call 340 IDs per hour.
We estimate the costs of running this measurement on a cloud computing
platform such as EC2 \cite{EC2} to be approximately \$500 per week.

Preliminary tests suggest that it would have been possible to increase
the number of called users by one order of magnitude with
virtualization.  Indeed, the main issue we faced is that running
several tracking clients on a machine makes it harder to isolate
packets from each client. One solution we tested but did not use in
our scheme, is to run several tracking client per physical machine,
each client in a different virtual machine.  Because the goal in this
paper is to demonstrate the feasibility of our scheme and not to fully
optimize it, running a single tracking client per machine is
sufficient.

\subsubsection{Measurement Results}
\begin{figure*}[!ht]
  \centering
  \includegraphics[width=0.62\columnwidth]{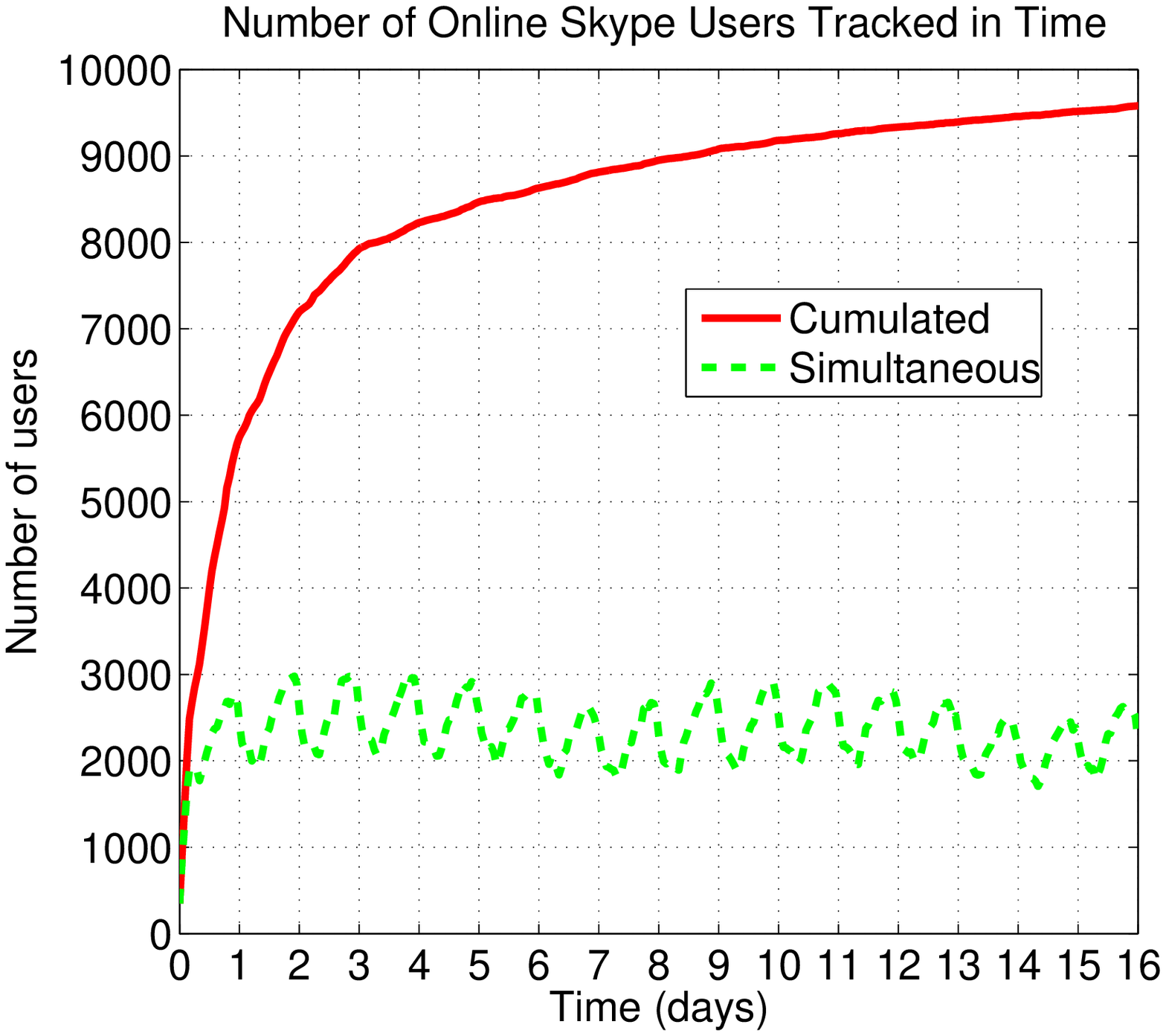}
  \includegraphics[width=0.65\columnwidth]{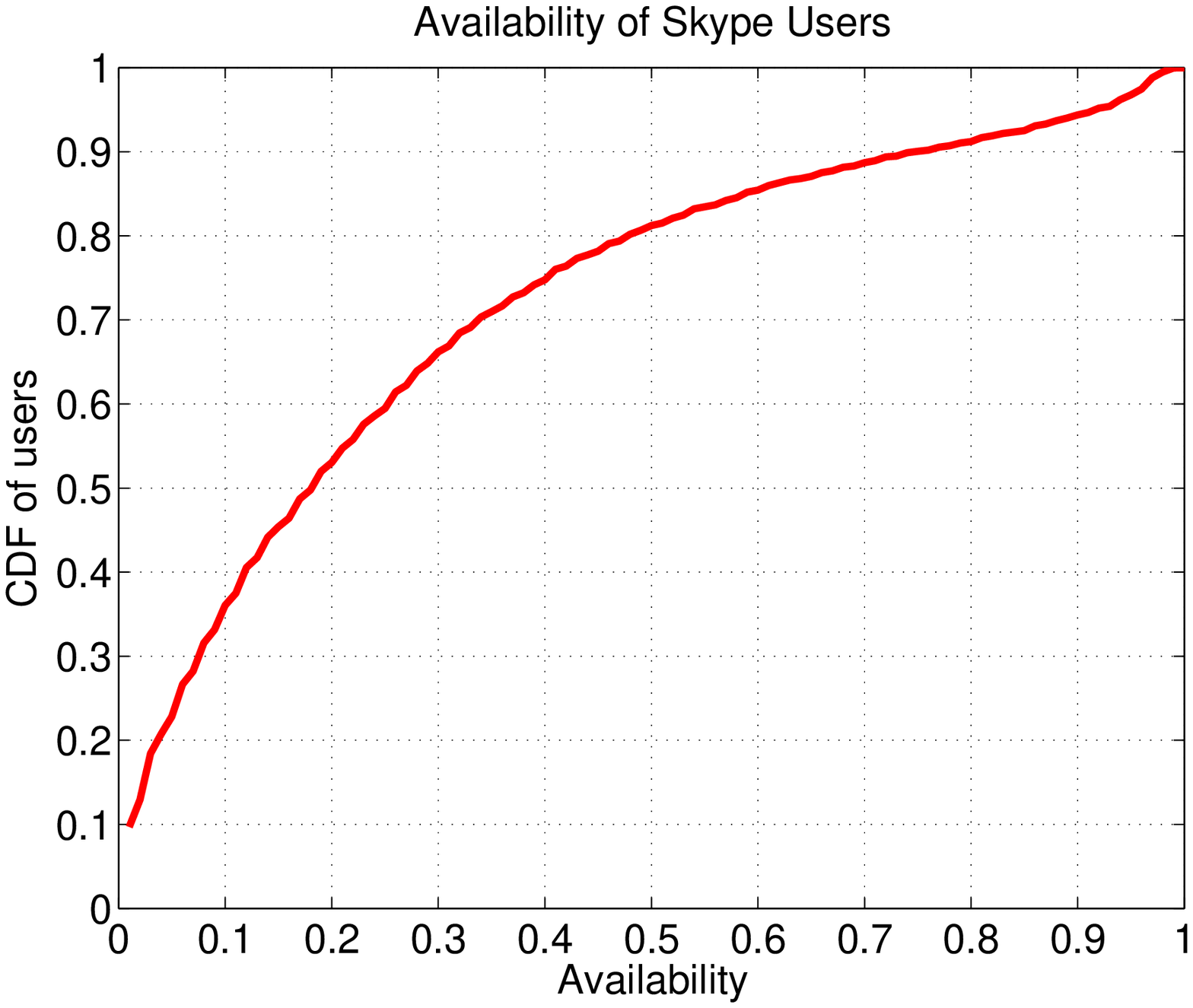}
  \includegraphics[width=0.60\columnwidth]{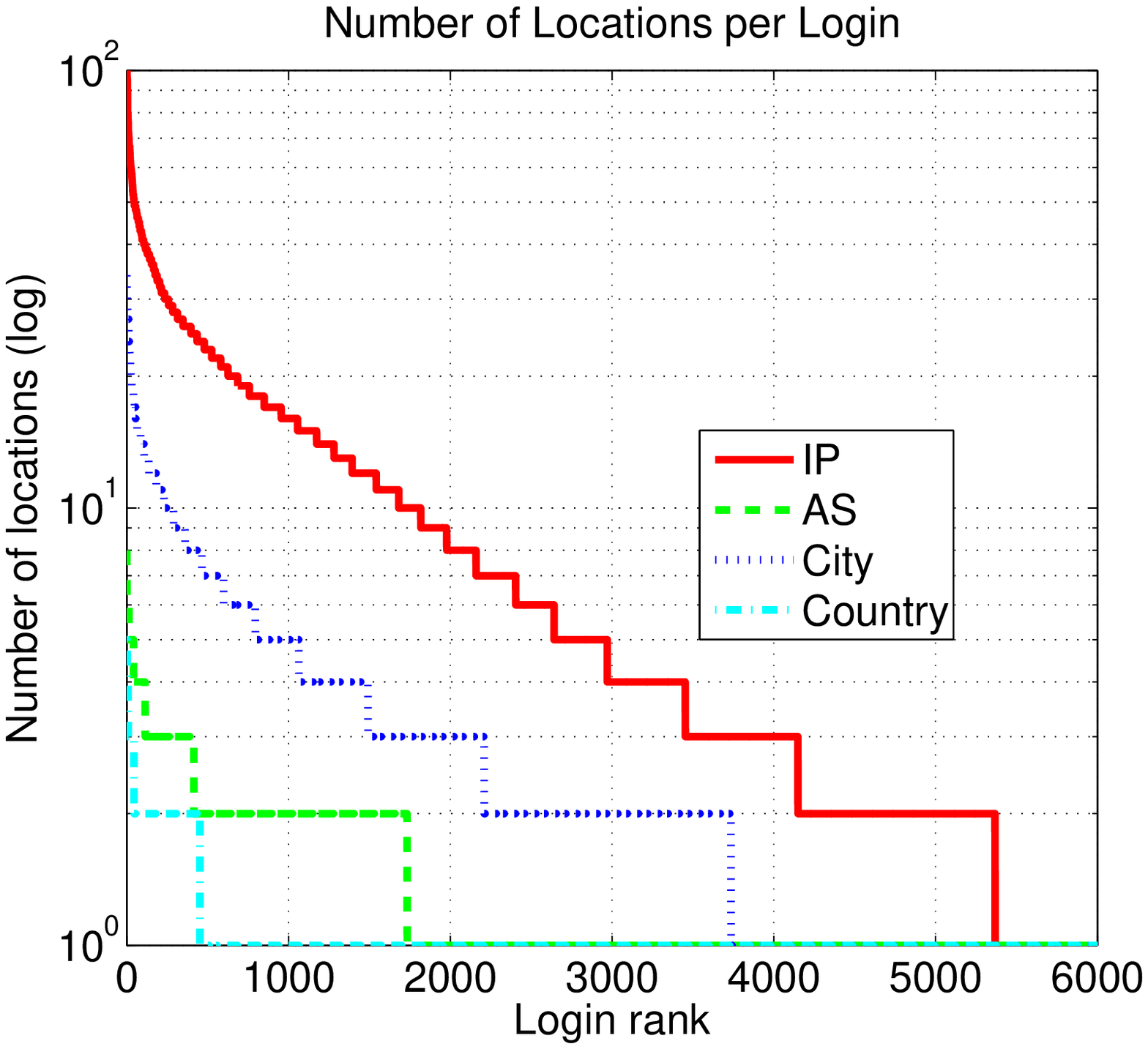}
  \caption{(left) Number of simultaneous and cumulative unique online
    Skype users (of 10,000) called in two weeks.  (middle) CDF of
    availability of Skype users. (right) Number of locations visited
    in two weeks by each Skype user, sorted by decreasing number of
    locations.  \textit{Skype users are mobile.}}
  \label{fig:location-online}
\end{figure*}

Whether our scalable calling scheme actually captures the {\it
  mobility} of a significant fraction of Skype users depends on three
questions that we address in the following.

{\it 1) Is it possible to periodically call a large number of Skype
  users?}  In Fig.~\ref{fig:location-online} (left), we see that at any
given time, we are calling between $2,000$ and $3,000$ online users
among the 10,000 users.  The diurnal behavior is due to the
heterogeneous distribution of Skype users worldwide. A large fraction
of Skype users are from the US and Western Europe. So during the
daytime in the US and Western Europe, there are more Skype users
online than during night in these geographical areas.  We also see in
Fig.~\ref{fig:location-online} (left) that after two weeks, we have
found at least one current IP address for $9,500$ users, which
represents $95$\% of the users we were periodically calling.  In
summary, it is possible to periodically call a large number of Skype
users.

{\it 2) How often are Skype users online?}  We define availability as
the fraction of the time a given user is online.  In
Fig.~\ref{fig:location-online} (middle), we plot the CDF of
availability for the 9,500 Skype users that we have seen online at
least once.  Skype users are surprisingly available with 20\% of all
users available more than 50\% of the time.  One explanation for this
behavior is that the Skype clients starts automatically at the startup
of the system.  In summary, Skype users are highly available so one
can call them to collect their location most of the time.

{\it 3) Can Skype users be found in several locations?}  Mobility
results in a change of IP address geo-localized in a different city,
AS, and/or country.  For each user that is online at least once, we
determine the different locations he visits over the two-week period.
This location information is anonymized (see
section~\ref{sec:single-geo}). In Fig.~\ref{fig:location-online}
(right), we see that 40\% of the 9,500 Skype users change city, 19\%
change AS, and 4\% change country at least once in two weeks.  In
summary, Skype users run Skype from several locations so one can
observe their mobility.  In summary, Skype users often run Skype from
different locations, and this mobility can be tracked by our
methodology.

Our methodology to measure the number of locations of a user has two
limitations.  First, in some cases (e.g., dynamic IP address), MaxMind
might erroneously associate a same user to different locations.  We
believe that such errors are very unlikely at the scale of a country
or an AS, and only occurs rarely at the scale of cities so that it
does not significantly impact our conclusions (see
Section~\ref{sec:single-geo}).  Second, the IP address may not capture
the location of users running Skype on their mobile phones
\cite{3G-phone}.  Although this may impact our ability to track Skype
users in the future, we believe that relatively few users fall into
this category today.

We may observe that significantly more users are mobile among cities
than among ASes for two reasons.  First, some ISPs have broad
geographical coverage, so users located in those ISPs are likely to
move within the same ISP, even though they change city.  Second, some
ISPs provide country-wide free Wifi hotspots to their users.  When
users of such ISPs change of city, they are likely to use these
hotspots, thus connecting from the same AS but a different city.

We note that, as the accuracy of IP geo-localization improves, it will
be possible to determine the locations of users with much finer
granularity.  For instance, a recent paper shows that it is possible
to geo-localize IP addresses with a median error distance of 690
meters \cite{Geolocalization}.

\section{File-Sharing Usage of Skype Users}
\label{sec:filesharing}

In the previous sections, we established that it is possible to map a
person to his IP address in a scalable manner.  We are now interested
in validating that this scheme introduces a linkability threat where
the identity of a person can be associated to his Internet usage.  In
particular, we focus in this section on finding the identity of
file-sharing users.  We focus on the BitTorrent application; however,
other P2P applications -- such as eMule \cite{emule} or Xunlei
\cite{xunlei} -- could instead be used.

One of the challenges here is that many file-sharing users are NATed,
that is, they may share their IP address with several users. We
present in the following a scheme exploiting the identification field
in the IP datagrams to check whether two different applications
actually run on the same machine. To the best of our knowledge, we are
the first ones to run and validate such a scheme in the wild.

In this section, we anonymized (as described in
section~\ref{sec:single-geo}) all localization information, we do not
store IP addresses after the verification procedure, and we never
store any information (including the infohash and the content name)
related to the contents downloaded by a given user.

\subsection{Methodology}

Our measurement system comprises a \textit{Skype tracker}, an
\textit{Infohash crawler}, a \textit{BitTorrent crawler}, and a
\textit{Verifier} which communicate through shared storage.  We begin
by randomly selecting a set of 100,000 identified Skype users.  The
Skype tracker employs ten tracking clients to daily collect the IP
address for the 100,000 users.  The Infohash crawler determines the
infohashes (file identifiers) of the 50,000 most popular BitTorrent
swarms.  Operating in parallel with the Skype tracker, the BitTorrent
crawler collects the IP addresses participating in the 50,000 most
popular swarms, and determines the IP addresses found in both Skype
and BitTorrent.  Finally, the Verifier attempts to initiate P2P
communications with the two applications in order to verify that the
same user is indeed running both of them.  In the following, we
describe in more detail the operation of each component.  The
operation of the Verifier will be described in
Section~\ref{sec:filesharing-verifier}.

\paragraph{The Skype Tracker} We use the methodology developed in
Section~\ref{sec:mapping} and Section~\ref{sec:location} to find
100,000 active Skype users.  In order to daily call 100,000 Skype
users, the Skype tracker uses ten tracking clients.  Because we are
now not interested in fine grain mobility measures but instead in
file-sharing usage, we only call each user once per day.  We then
analyze packet patterns to determine the latest IP address of these
users and temporarily save them to a shared storage.  (Keep in mind we
collect the IP addresses not only of users that are online but also of
all users that have logged into the system in the last 72 hours.)
These IP addresses are then loaded from the shared storage by the
BitTorrent crawler to determine which files are distributed from these
IP addresses.

\paragraph{The Infohash Crawler} We collect file identifiers
(infohashes) from the PublicBitTorrent tracker \cite{public-BT}, which
is the largest BitTorrent tracker at the time of this writing.
PublicBitTorrent publishes a file with all the infohashes it tracks on
its website.  This file is the dump of a request, \textit{scrape-all},
supported by trackers running the OpenTracker software
\cite{BT-Spying}.  This request returns all infohashes of files it is
tracking and the number of downloaders (leechers) and uploaders
(seeds).  We download this file every day from the PublicBitTorrent
website and extract the infohashes for the 50,000 most popular files.

\paragraph{The BitTorrent Crawler} 
In this step, we seek an efficient mechanism to obtain the IP
addresses participating in the 50,000 most popular torrents.
BitTorrent trackers such as PublicBitTorrent support a request,
\textit{announce started}, that returns a list of peers participating
in a torrent identified by an infohash.  As tracker developers became
aware that such requests can be abused they started to limit the
number of requests a given peer can send before being blacklisted.
Therefore, instead of using the PublicBitTorrent tracker to collect IP
addresses, we use a decentralized tracker (DHT).

We collect the IP addresses participating in the top 50,000 torrents
from the Mainline DHT every hour for two weeks.  This DHT is a
decentralized tracker that is primarily used by $\mu$Torrent
\cite{uTorrent} and Mainline BitTorrent \cite{bittorrent}, the most
popular BitTorrent clients.  However, we note that other popular P2P
file-sharing clients, such as Xunlei, also support it.

When a peer wants to download a new file, it contacts the Mainline DHT
to obtain a list of peers distributing that file.  This peer first
finds the DHT node maintaining the list of peers for that file using
the \textit{find\_node} request.  That request takes an infohash as a
parameter, and essentially returns the ID and (IP, port) pair of the
DHT node responsible for that infohash.  Then, the peer sends a
\textit{get\_peers} request to that node, which returns a list of (IP,
port) pairs belonging to peers distributing the file.

Unlike centralized trackers, we observed that DHT nodes do not
implement blacklisting strategies.  So we located the nodes
responsible for the 50,000 files that we wanted to crawl and then
repeatedly sent get\_peers requests to collect the peers distributing
these files.  The whole procedure distributed over 10 machines takes
about one hour.

Each of our crawling bots periodically loads the (Skype\_ID, IP) pair
of active Skype users into memory.  If the IP address of an active
Skype user is also found in a BitTorrent swarm, the user is possibly
downloading the corresponding file (this correlation is performed
on-the-fly and we never store the mapping IP address, infohash).  {\em
  However, we must verify this hypothesis as an IP address may
  correspond to a NAT shared by several users.}  We refer to this
problem as the NAT problem.  We note that several types of
middleboxes, including NATs and IPv6 routers can use a single public
IP address for different users. For the sake of simplicity, we use the
term NATs when we refer to the generic notion of middleboxes in the
following.  (We note that dynamic IP addresses can also be shared by
several users, resulting in the same problem.)

\subsection{The NAT Problem}

\begin{figure}[!t]
  \centering
  \includegraphics[width=0.962\columnwidth]{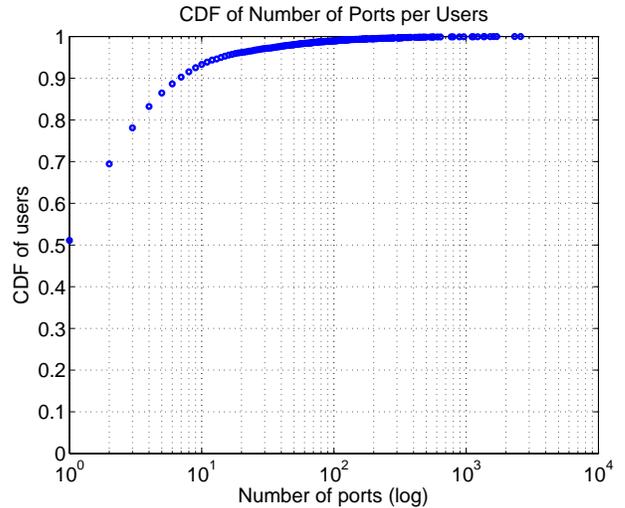}
    \caption{CDF of the users using BitTorrent as a function of
      BitTorrent ports. \textit{50\% of collected BitTorrent users
        share their IP address with other BitTorrent users.}}
            \label{fig:usage-NAT}
\end{figure}

Depending on the Internet connectivity of a user, an IP address may
correspond to a computer or to a NAT shared by a household, a company,
or even an ISP.  Because several users can share the same IP address,
we may wrongly associate an identified Skype user to the BitTorrent
downloads of another user behind the same NAT.  To the best of our
knowledge, all BitTorrent clients multiplex torrents on a single port.
This port is picked at random at the installation of the client, and
remains the same in subsequent utilizations.  Therefore, we can
associate each IP/port pair to a single BitTorrent user
\cite{BT-Spying}.  However, this observation alone does not allow us
to match a Skype user to a BitTorrent user when the user is behind a
NAT, as described below.

We found 15,000 users (out of 100,000) who have IP addresses that were
simultaneously found in Skype and BitTorrent during a period of two
weeks.  Of these 15,000 Skype users using BitTorrent, approximately
7,500 (50\%) share their IP address with another BitTorrent user (as
indicated by users with more than one port in
Fig.~\ref{fig:usage-NAT}).  In other words, a significant fraction of
the 15,000 Skype users are behind a NAT and may therefore not be the
ones using BitTorrent (false positives).

\subsection{The Verifier}
\label{sec:filesharing-verifier}

We now describe the operation of our Verifier tool, which is
responsible for definitively establishing whether Skype and BitTorrent
are run on the same machine.  Although more than one person
\textit{simultaneously} share the same machine, the granularity of a
machine is enough for our purpose.  For the sake of simplicity, we
assume in the following that each machine is used by a single person.

Given an IP address that participates in both Skype and BitTorrent
(matching IP), we now describe how the Verifier makes sure the person
identified in Skype is indeed the one using BitTorrent.  Consider a
scenario where two users, Alice and Bob, are behind the same NAT.
Suppose that, by calling Alice on Skype, we have determined that her
IP address is in a swarm in BitTorrent, but the IP address is a NATed
one.  Two scenarios are possible.  In the first scenario, Alice is
using both Skype and BitTorrent on the same host.  In the second
scenario, Alice is using Skype on one host and Bob is using BitTorrent
on another host.  The second scenario corresponds to a false positive
because Alice is not the one using BitTorrent.

To detect false positives, we leverage the predictability of the
identification field in the IP datagrams (IP-ID) originating from the
same machine \cite{Counting}.  As soon as the BitTorrent crawler
detects a matching IP address, it signals the Verifier, which
immediately calls the corresponding Skype user and, at the same time,
initiates a handshake with the BitTorrent client.  If the distance
between the IP-IDs generated by Skype and those generated by
BitTorrent is small, Alice is very likely to be the identified
BitTorrent downloader.  Otherwise, Alice is likely to be a false
positive.

At the end of the verification procedure, IP addresses are anonymized
using a salted hash. All subsequent analysis is performed on this
anonymized data. 

\paragraph{Limitations} Our verification procedure has two
limitations.  The first limitation is that we can only initiate
communication to public peers or NATed peers that accept incoming
communications (e.g., when UPnP is used).  This limitation
significantly restricts the number of BitTorrent users we can verify.
However, for this proof of concept, it is not necessary to verify all
the Skype users who are downloading with BitTorrent.  An aggressive
attacker could easily verify more users by registering the IP address
of the Verifier to the Mainline DHT. In this manner he would also
receive incoming communication from peers whose NATs refuse incoming
communications.  Therefore, an attacker could in principle verify
NATed peers also.

The second limitation is that we assume that the IP-IDs originating
from the same machine are predictable, which depends on two
conditions.  The first condition is that the IP-IDs originating from
the same machine should be predictable (e.g., sequential).  Because
IP-IDs are attributed by the TCP stack of an Operating System (OS),
this first condition highly depends on the fraction of OSes observed
in the wild whose attribution is indeed predictable.  By testing
Windows XP, Vista, and 7, we verified that they all use sequential
IP-IDs.  As these three versions of Windows alone account for 90\% of
all OSes found in the wild \cite{OS}, we conclude that this first
condition is largely met.  The second condition is that NATs do not
modify the IP-IDs as attributed by the TCP stack of the machine.  This
condition is supported by $(i)$ related work in which this behavior
was not observed in practice \cite{Counting} and by $(ii)$ the
specification of the IPv4 ID field, which specifies that NATs should
ignore this field \cite{IP-ID}.

In conclusion, we expect that our verification procedure based on the
predictability of the IP-ID field to be highly accurate, that is, with
no, or few, false positives (due to similar IP-IDs originating from
different machines) and relatively few false negatives (due to OSes
with unpredictable IP-IDs attribution or IPv6 routers that
re-attribute IP-IDs unpredictably).

\subsection{Experimental Results}  

By running our verification procedure for two weeks, we successfully
triggered communication between the Verifier and 765 unique users on
both Skype and BitTorrent.  We refer to these users as {\em
  verifiable}.

\begin{figure}[!t]
  \centering
  \includegraphics[width=0.962\columnwidth]{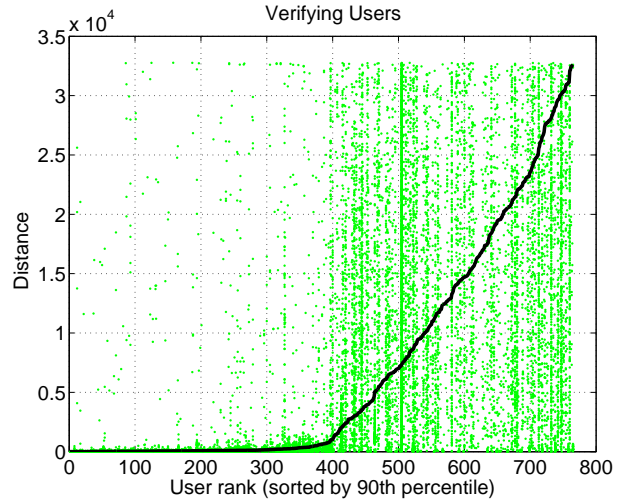}
  \caption{After two weeks, we plot the 90th percentile of the
    shortest distance between the IP-IDs on a ring of $2^{16}$
    elements of the first Skype and BitTorrent packets received from a
    verifiable user, sorted by increasing 90th percentile (curve).
    There is one dot per verification experiment.  \textit{We verify
      400 users out of 765 users.}}
    \label{fig:usage-verification}
\end{figure}

We investigate the fraction of verifiable users that we actually fully
verified.  For the 765 verifiable users, we compute the shortest
distance on a ring of $2^{16}$ elements between the IP-IDs of the
first packet received from Skype and from BitTorrent.  The smaller the
distance, the more likely the identified Skype user is indeed using
BitTorrent.  In Fig.~\ref{fig:usage-verification}, we see that running
this procedure finds 400 unique users for whom the 90th percentile of
the distance is less than 1,000.  We conclude that approximately 400
users (52\% of the 765 verifiable users) are indeed using BitTorrent.
We cannot conclude for sure that the remaining 48\% of the verifiable
users are not BitTorrent users (they might be false negatives).
However, as we have seen that at least 90\% of the OSes use sequential
IP-IDs, we strongly believe that most of them are not using
BitTorrent.

\begin{table}
  \small
  \centering
  \begin{tabular}{|c|c|c|c|c|c|}
    \hline Rank & \# Files & First name & Last name &  City &
    Country \\ \hline

    1 & 23 & \ding{51} & \ding{51} & \ding{51} & \ding{51}\\
    2 & 18 & \ding{51} & \ding{51} & \ding{51} &  \ding{51}\\
    3 & 12 & \ding{51} & \ding{51} &  \ding{55} &  \ding{51}\\
    4 & 11 & \ding{51} & \ding{51} & \ding{51} &  \ding{51}\\
    5 & 11 & \ding{51} & \ding{51} & \ding{51} &  \ding{51}\\
    6 & 11 & \ding{51} & \ding{51} & \ding{51} &  \ding{51}\\
    7 & 9 & \ding{55} & \ding{51} & \ding{51} &  \ding{51}\\
    8 & 8 & \ding{55} & \ding{51} & \ding{51} &  \ding{51}\\
    9 & 7 & \ding{51} & \ding{51} &  \ding{51} &  \ding{51}\\
    10 & 6 & \ding{51} & \ding{51} &  \ding{51} &  \ding{51}\\

    \hline
  \end{tabular}
  \caption{For each of the top10 verified user, we show the number of
    files shared by that user, whether the user provides in its Skype
    profile a first or last
    name, a city, and a country.}
\label{tab:usage-top10}
\end{table}

In summary, we have determined 400 identified Skype users (from a
random set of 100,000) who are definitely using BitTorrent.
Table~\ref{tab:usage-top10} shows the information that is readily
available about the top-10 BitTorrent users.  When registering with
Skype, all of these users provided their last names and all but two
users also provided their first names. In addition, all but one of
these users provided their cities of residence. However, we remind
that we do not store any personal information (e.g., name and city)
for the purpose of this measurement; instead, we only store a binary
information indicating whether a personal information is available or
not.

\section{Defenses}
\label{sec:defenses}

In the previous sections we have seen that it is possible for an
attacker to develop and deploy (possibly from a home) a tool that
periodically determines the current IP address of a targeted VoIP
user. Even if the VoIP user is behind a NAT, the attacker can
determine the user's public IP address. Observing the mobility of a
targeted individual could be used for many malicious purposes.  In
this section we briefly discuss defenses for this attack, both at the
application level and at the user level.

One measure that can go a long way is for the designers of the VoIP
signaling protocol to simply {\em ensure that the callee's IP address
  is not revealed to the caller until the callee accepts the
  call}. That is, before the callee accepts the call, callee's
signaling packets are sent to supernodes or infrastructure nodes, and
not to the caller; furthermore, the caller is not provided the
callee's IP address during call set-up. By only revealing the callee's
IP address after the callee accepts the call, then $(i)$ it is no
longer possible to make an inconspicuous call to the target; and
$(ii)$ if Alice chooses to block all calls from strangers (i.e.,
people not on her contact list), then a stranger will no longer be
able to determine her IP address and observe her mobility. This
solution has a very low overhead as only a few signalling messages are
relayed. Thus, we strongly recommend that all VoIP applications adopt
this simple mechanism.

However, even with this simple mechanism in place, a friend of Alice
(that is, anyone on her contact list, including friends, old
boyfriends, family members, employers, and employees) would still be
able to determine her IP address (and location) when they call her and
she accepts the call. We now outline some measures that defend against
this attack.

One blanket defense for these attacks is to have all calls pass
through relays. When a datagram passes through a relay, the relay
regenerates the datagram with the source IP address of the relay. If
the relay can be trusted, then neither party in the call sees the
other's IP. In fact, in Skype, if both caller and callee are behind a
NAT, then the call is typically relayed through a third skype user
(who is not behind a NAT), serving as a relay.  The relays must be
selected so as not to give away the location of the callee.  (For
example, the system shouldn't strive to find a relay in same city as
the callee.)  The main problem with this solution is that it detracts
from the efficiencies of P2P communication because $(i)$ relays must
now be made available to support the huge bandwidth demands of
large-scale real-time voice and video communication systems; and
$(ii)$ access ISPs will see an increase of upstream and downstream
relay traffic.

In order to not excessively route traffic through relays, the system
can be designed so that Alice can specify for which contacts in her
address book the calls are to be routed through relays. For example,
if Alice is only concerned about her boss observe her mobility, she
can configure her client to have calls between her and her boss pass
through relays. The client could also be designed to make this
decision on a call-by-call basis: whenever, her boss attempts to call
her, she is asked whether this should be a P2P or relayed call.

We briefly mention that another approach for providing location
privacy is to run the P2P communication application through a
third-party anonymizing service such as Tor \cite{Tor}.  However, the
delay and throughput performance of Tor and similar services is
clearly insufficient for supporting real-time voice and video
\cite{oneswarm, Waiting-for-anonymity}.  In addition to being
inefficient, Tor also introduces privacy issues for certain
applications (e.g., P2P file sharing) \cite{Bad-Apple}.

We conclude this brief discussion on defenses by mentioning that these
location attacks actually have their roots in the current Internet
architecture, for which all datagrams carry source and destination IP
addresses. We are not advocating a total re-design of the Internet,
but we mention that this and other Internet privacy problems could be
resolved by using alternative underlying network architectures. For
example, if the Internet were to use virtual circuits (as with X.25
and ATM), then it would be much more difficult for a stranger or a
friend to observe a user's mobility.

\section{Related Work}
\label{sec:related}

\subsection{Mobility}

We now describe the related work on observing the mobility of users by
using IP addresses and cell phones.

\paragraph{IP Address Mobility} Guha et al. \cite{dns} is the work on
IP address mobility that is the closest to ours.  The authors show
that by periodically retrieving the IP address of dynamic DNS users,
an attacker can observe the mobility of these users.  Whereas the goal
of their attack is similar to ours, there are two major differences
between exploiting dynamic DNS and Skype to measure mobility.  First,
dynamic DNS allows to infer the identify of the user in ``some cases''
whereas we have showed that 88\% of Skype users provide their birth
name, and that 82\% also provide either age, gender, homepage,
country, or language.  Second, targeting dynamic DNS users limits the
scope of the attack.  Whereas there are a few millions users of
dynamic DNS in the world, we showed that much more Skype users are
susceptible to have their mobility tracked.

\paragraph{Cell Phone Mobility} The Carmen Sandiego Project
\cite{Carmen-Sandiego} recently showed how to use cell phones to
observe the mobility of a user.  The authors first use the caller ID
service to collect persons-to-cell phone numbers mappings.  Then, by
accessing the Home Location Register (HLR), they show that an attacker
can collect the current Mobile Switching Center (MSC) identifier for a
given phone number.  As MSC identifiers often gives the indication of
the location of a user, an attacker can periodically collect that
information to observe the mobility of an identified cell phone user.
One important weakness of this attack is that there is no convention
on how an operator attributes MSC identifiers.  So the naming
convention for MSCs varies from one operator to the other and it is
hard to determine to which location a given identifier corresponds.

Even though it is not our primary purpose, we believe our scheme, and
in particular the description of Skype packet patterns between caller
and callee, also has the potential to significantly simplify the
tracing of Skype calls.

To the best of our knowledge, we are the first to show that it is
possible to use real-time applications to map a person to an IP
address and to scale that scheme to observe the mobility of a large
number of persons.  As we have shown it might be possible to observe
the mobility of 56 million identified Skype users worldwide at any
moment in time, we claim that the scope and the severity of our attack
are very severe.

\subsection{File-sharing Usage}

We now describe the related work on observing file-sharing usage and
verifying users.  Because we have used BitTorrent in this paper and it
is one of the most popular file-sharing system, we focus on BitTorrent
in the following.  However, we remind that all file-sharing systems
are in principle vulnerable to our attack.

In recent works, the scale of BitTorrent measurements has
significantly increased \cite{BT-Public-Ecosystem, BT-Monitors,
  BT-Spying}.  For example, Zhang et al. collected 5 million IP
addresses in 12 hours \cite{BT-Public-Ecosystem}, Siganos et
al. collected 37 million IPs in 45 days \cite{BT-Monitors}, and Le
Blond et al. collected 148 million IPs in 103 days \cite{BT-Spying}.
As noted by Le Blond et al. and more recently by Wolchok et
al. \cite{Crawling-BitTorrent}, being able to continuously collect the
IP addresses is a serious privacy threat in itself.  In this paper
though, we have not only collected the IP addresses of a large number
of BitTorrent users but we have also identified a significant fraction
of these users.

A security threat noted by Piatek et al. consists in injecting the IP
address of random Internet users into BitTorrent trackers to falsely
implicate them into copyright infringement \cite{DMCA}.  We note that
the ability to map a targeted person to an IP address significantly
worsens this threat because an attacker could also implicate that
particular person into copyright infringement.

As far as we know, we are the first to show that it is possible to
find the identity of BitTorrent users without requesting that
information from an ISP.  We believe that this attack introduces a
serious potential for blackmail and phishing attacks.

\paragraph{Verification} We relied on IP-IDs to verify the identity of
BitTorrent downloaders.  This technique has been used in the context
of passively counting the number of machines behind a NAT
\cite{Counting} (on a LAN).  As far as we know, it has never been used
on the Internet to actively verify that several applications were
running on the same machine.  Alternatively, we could have used remote
physical device fingerprinting \cite{Device-Fingerprinting} but using
IP-IDs was simpler and sufficient for our purpose.

\section{Conclusion}
\label{sec:conclusion}

We have shown that it is possible for an attacker, with modest
resources, to determine the current IP address of identified and
targeted Skype user (if the user is currently active). It may be
possible to do this for other real-time communication applications
that also send datagrams directly between caller and callee (such as
MSN Live, QQ, and Google Talk).  In the case of Skype, even if the
targeted user is behind a NAT, the attacker can determine the user's
public IP address.  Such an attack could be used for many malicious
purposes, including observing a person's mobility or linking the
identity of a person to his Internet usage.

We have further shown that by deploying modest resources, it is
possible for an attacker to scale this scheme to not just one user but
tens of thousands of users simultaneously. A prankster could use this
scalable calling scheme to, for example, create a public web site
which provides the mobility and file-sharing history of all active
Skype users in a city or a country.  Parents, employers, and spouses
could then search such a web site to determine the mobility and
file-sharing history of arbitrary Skype users.

\bigskip

{\bf Acknowledgements.} We would like to thank the volunteers for the
considerable amount of time they have spent to help us test our
scheme.  We also would like to thank Justin Cappos, David Choffnes,
Paul Francis, Krishna Gummadi, Engin Kirda, and the anonymous
reviewers for their constructive comments.  This research was
partially supported by the NSF grant CNS-0917767.

\bibliographystyle{acm}
\bibliography{biblio}

\end{document}